\documentclass[aps,prl,twocolumn,superscriptaddress,nofootbib]{revtex4-2}

\usepackage{amsmath,amssymb,bm}
\usepackage{graphicx}
\usepackage{hyperref}
\usepackage{cleveref}
\usepackage{xcolor}

\newcommand{\dd}{\mathrm{d}}

\begin{document}

\title{The Double-Copy Root of Hawking Thermality}

\author{John Joseph M. Carrasco}
\affiliation{Amplitudes and Insights Group, Department of Physics and Astronomy, Northwestern University, Evanston, IL 60208, USA}
\affiliation{Center for Interdisciplinary Exploration and Research in Astrophysics (CIERA), Northwestern University, 1800 Sherman Ave, Evanston, IL 60201, USA}

\author{Yaxi Chen}
\affiliation{Amplitudes and Insights Group, Department of Physics and Astronomy, Northwestern University, Evanston, IL 60208, USA}
\affiliation{Center for Interdisciplinary Exploration and Research in Astrophysics (CIERA), Northwestern University, 1800 Sherman Ave, Evanston, IL 60201, USA}

\date{\today}

\begin{abstract}
The Hawking radiation spectrum from a collapsing null shell can be derived via the double copy of a simpler gauge theory calculation. Analyzing the non-abelian Yang-Mills root of this process, we demonstrate that the radiation spectrum is thermal in the color charge eigenvalue $\lambda$, not energy. Considering the $SU(N_c)$ gauge theory in the large $N_c$ limit, we find the differential spectrum $\dd N/\dd \lambda$ is a product of the gravitationally familiar Planck-like factor and the color phase space density, modeled here as the Wigner semicircle from random matrix theory. This reveals that apparent energy thermality in gravity is the direct dual of charge thermality in its underlying non-abelian gauge theory.
\end{abstract}

\maketitle

\section{Introduction}
Classical solutions in General Relativity (GR), such as the Schwarzschild metric, can be viewed as an infinite sum of tree-level graviton diagrams sourced by their own self-interaction \cite{Duff:1973zz}. This perturbative picture suggests an immense complexity that is miraculously resummed in the exact solution. The classical double copy~\cite{Monteiro:2014cda, Luna:2015paa,White:2024pve} reveals this is no miracle, but a direct consequence of gravity's structure as the relativistic \emph{quantum double copy}~\cite{Bern:2008qj, Bern:2010ue, Bern:2019prr,Adamo:2022dcm} of a non-abelian Yang-Mills (YM) theory.

This principle is clearly manifest in metrics of the Kerr-Schild form, $g_{\mu\nu} = \eta_{\mu\nu} + \phi_g k_\mu k_\nu$. For a null, geodesic vector $k_\mu$, the Einstein tensor linearizes, $G_{\mu\nu}[g] = G^{(1)}_{\mu\nu}[h]$, meaning the full non-linear field equations reduce to a free wave equation. The infinite tower of graviton interactions collapses, and the solution behaves as if generated by a single, simple propagator. This simplicity is a direct consequence of the double copy, which guarantees that every operator in the Einstein-Hilbert action is completely specified by YM operator data~\cite{Bern:1999ji,Carrasco:2025ymt}. 

% "This ain't no miracle, it's math bitches." --- JAK
This correspondence is realized in the YM root. There, a specific gauge choice allows a potential with a fixed color orientation, $A^a_\mu = c^a \phi k_\mu$, to also behave as a simple propagator; it becomes an exact classical solution because its non-linear self-interactions vanish identically via the color algebra ($f^{a_1 a_2 a_3}c^{a_2} c^{a_3} = 0$ if $c$ is aligned along a fixed color direction). The Kerr-Schild metric is therefore not merely analogous to a double copy; it is the literal double copy of the Yang-Mills propagator in the presence of an abelianized source. The geometric properties that linearize Einstein's equations are the precise dual of the gauge choice that reduces Yang-Mills to its fundamental propagator.

While the classical solution appears abelian, the underlying root theory of general relativity must be non-abelian Yang-Mills. The duality relies on the kinematic numerators of YM amplitudes; indeed, their cubic and quartic gauge self-interactions are sufficient to generate all operators in GR through the double copy. Attempting to double-copy QED yields linearized gravity only. It has a consistent metric interpretation but none of the self-interaction that is the hallmark of Einstein-Hilbert. The celebrated generation of Newton from Coulomb is possible only because their non-relativistic potentials are described solely by mediator exchange in two-to-two scattering. Einstein requires Yang-Mills.

In a remarkable paper~\cite{Aoude:2024sve} of this past year, Aoude, O'Connell, and Sergola recovered the apparently thermal spectrum of Hawking radiation by emphasizing the on-shell nature of the original~\cite{Hawking:1975vcx} calculation, setting the stage for a modern S-matrix perspective. The physical foundation of their approach lies in the principles of quantum field theory on a dynamic background. For a collapsing shell, the initial vacuum state ($|0_{\text{in}}\rangle$) is not an eigenstate of the final Hamiltonian and evolves under the S-matrix into a superposition of states containing real, outgoing particles --- a formal description of particle creation from vacuum fluctuations.

Ref.~\cite{Aoude:2024sve} reminds us that we can compute the spectrum of this created radiation by considering the evolution of a single probe state. This probe is a computational tool; we calculate its scattering amplitude to characterize its dynamical interaction with the background responsible for particle creation. The calculation begins by computing a three-point tree-level amplitude for a probe scattering against the Vaidya background. By exponentiating this result via the Lippmann-Schwinger equation, the authors capture the exact result within the eikonal limit~\footnote{The eikonal limit describes the dynamics of particles whose trajectories are nearly light-like and forward-scattered. These are precisely the kinematics of the modes that just skim the forming event horizon to become the observed Hawking radiation.}, which re-sums the leading soft contributions to all orders.

The crucial result of this calculation is a logarithmic eikonal phase, $\chi(v_0) \propto \log(-v_0)$. The argument, $v_0$, represents the probe trajectory's initial time offset $v_0<0$ relative to the moment of collapse. This phase is not specific to the probe but is a universal imprint left on \textit{any} quantum mode by the extreme time-delay near the forming horizon. It is this mathematical structure that encodes the mixing of positive and negative frequency modes (a Bogoliubov transformation) that defines particle creation. The spectrum derived from the probe's phase shift is therefore the spectrum of the particles spontaneously created from the vacuum. Critically, the Kerr-Schild nature of the Vaidya metric provides a unique opportunity to study this process not through the weak-strong duality of holography, but through the weak-weak duality of the double copy.

\emph{Note added in preparation:} During the final stages of preparing this letter, two papers~\cite{Aoude:2025jvt,Ilderton:2025aql} appeared with coordinated release. These papers carefully calculate in the abelian (Maxwell) limit of this root-Vaidya setup, confirming it yields a non-thermal, Bremsstrahlung-like energy spectrum. Our work complements their analysis by considering the non-abelian YM root. We find that the apparent thermality hidden in the abelian energy spectrum re-emerges in the natural charge of the full root theory: color.

\section{Derivation of the Non-Abelian Phase}
Here we summarize the derivation of the radiation spectrum from the eikonal S-matrix, referring readers to \cite{Aoude:2024sve} for a comprehensive treatment in the double-copy case.   The eikonal S-matrix describes the scattering of a probe of momentum $p$, initial energy $E=|\vec{p}|$, passing through the classical background in the limit of high-energy, low-momentum-transfer ($q=p'-p \to 0$) scattering. 

We consider the non-abelian root of the Vaidya metric: a shell of collapsing color charge $Q$ activating at light-cone time $v=t+r=0$. The gauge field is given by:
\begin{equation}
A^a_\mu(x)= c^a \frac{\theta(v) Q}{r} k_\mu.
\end{equation}
The null vector $k_\mu$ is chosen such that $k_\mu \dd x^\mu = \dd v$; in these coordinates, $k^2=0$. A critical consequence of this null property is that any four-point ``seagull'' interaction involving two insertions of the background field is proportional to $A_\mu A^\mu \propto k^2 = 0$. The background therefore does not source such contact terms; the probe couples entirely through the single-insertion (three-point) vertex.

We consider the evolution of a probe particle in initial wavepacket $|\psi\rangle$ centered on momentum $p$.  The  S-matrix element is the overlap between  $|\psi\rangle$   and the final on-shell state $|p'\rangle$.   
We introduce the temporal offset of the trajectory relative to shell formation via a timelike ``impact parameter'' $b^\mu_0 = (v_0,\vec{0})$.  At leading order, the eikonal S-matrix is an integral over the momentum transfer $q$:
\begin{equation}
 \langle p' | \hat S-1 | \psi \rangle_\text{LO} \propto \int \dd^4 q \, \delta(2p'\!\cdot\!q) \, e^{-iq\cdot b_0} \, i\mathcal{A}_{3}(p \to p').
\label{eq:lo-smatrix}
\end{equation}
The function $\mathcal{A}_{3}(p \to p')$ encodes the tree-level background contribution to the scalar's dynamics which we can read off from the three-point on-shell amplitude as we will discuss. The delta function enforces the on-shell condition ($2p'\cdot q \approx 0$ in the eikonal limit). This constraint reduces the momentum-space integral to an integral along the classical, light-like worldline of the probe, $x^\mu(\sigma) = b_0^\mu + 2\, \sigma  \, p'^\mu$. In this limit, the incoming probe is sufficiently energetic that its trajectory is unaffected by the background potential; it follows a rigid, classical worldline. Consequently, the scattering process does not change the probe's momentum, but only rotates its internal color vector.  The full S-matrix in the eikonal limit is found by solving the Lippmann-Schwinger equation, which re-sums soft radiation into an exponentiated eikonal phase,  $S \approx e^{i\chi}$. 

Physically, this S-matrix describes the parallel transport of the probe's color frame along its classical path, given formally by the path-ordered Wilson line.  The color structure simplifies because the background has a fixed orientation $c^a$. Because the background's color vector $c^a$ is held constant along the probe's worldline, the color operator at every point is simply $c^a T^a$. This means the matrices at different points on the worldline commute, and the path-ordering becomes trivial.   For a probe prepared in an eigenstate of this operator, the color algebra is replaced by the scalar eigenvalue $\lambda$, effectively abelianizing the interaction along the trajectory.

The starting point of the actual calculation is then the on-shell tree-level amplitude for a scalar probe scattering off a single gluon,
\begin{equation}
\mathcal{A}_3^{abc}(p,p',q)
\;\propto\;
g\, f^{abc}\,(p+p')^\mu \varepsilon_\mu(q)\,,
\end{equation}
where $\varepsilon_\mu(q)$ is the gluon polarization. In the presence of a classical background, the gluon is not a propagating quantum degree of freedom. Instead, the vertex describes the probe's response to the external field. Operationally, this replaces the polarization vector with the Fourier mode of the background, $\varepsilon_\mu(q) \to \tilde{A}^a_\mu(q)$.  Because of the fixed color orientation of the background, the contraction $f^{abc}c^a$  reduces to the action of a generator $c^a T^a$ on the probe which reduces to $\lambda$.

Combining these elements, the abelianized background contribution becomes:
\begin{equation}
\mathcal{A}_{3}(p\to p') =g\, \lambda \,Q\ \int \dd^4 x \, e^{i q  x} \frac{ \theta(v)}{r} k \cdot (p+p').
\label{eq:bg_contribution}
\end{equation}

Now we can calculate the eikonal phase $\chi$ determined by the tree-level integral \cref{eq:lo-smatrix} of the potential $A_\mu \propto (Q/r) k_\mu$ contracted with the probe momentum $p^\mu$ along the trajectory.   The eikonal worldline parameterization, $x^\mu(\sigma) = b_0^\mu + 2\, \sigma  \, p'^\mu$, and our normalization gives coordinates $r(\sigma)=2E |\sigma|$ and $t(\sigma)=v_0 + 2E \sigma$.  The  integral  \cref{eq:bg_contribution} is non-zero only where the source is active $(v = t+r >0)$ which establishes a physical, $v_0$-dependent lower bound on the integration region, $\sigma_{\text{min}} = - v_0/(4 E)$.    Changing variables from $\sigma$ to $r$ therefore cancels
the explicit factors of $E$, leaving,
\begin{align}
\chi(v_0) &= g \lambda Q  \int_{\sigma_{\rm min}}^\infty \dd\sigma \,  \frac{2 (p \cdot k)}{r(\sigma)} \nonumber \\
&= g \lambda Q \int_{r_{\text{min}}}^{\infty} \frac{\dd r}{2 E} \, \frac{2 (p \cdot k)}{r} \nonumber \\
&=  -g \lambda Q \log(-v_0/\mu).
\label{eq:eikonal_phase}
\end{align}
Here the IR scale $\mu$ absorbs the arbitrary upper cutoff. The cancellation of the $p \cdot k \approx E$ factors is the defining feature of the gauge theory root; the phase $\beta$ depends linearly on the color eigenvalue $\lambda$ but is independent of the probe energy $E$. We thus identify the phase coefficient $\beta = C\lambda$, defining an effective coupling constant $C = g Q$.

The final radiation spectrum is obtained from the Fourier transform of this time-dependent phase factor with respect to the radiated energy. As energy transfer is negligible in the eikonal limit, we identify the radiated energy with the energy of the incoming probe, $E$:
\begin{equation}
\label{eq:fourier_transform}
\mathcal{A}(E) = \int_{-\infty}^0 \dd v_0 \, e^{iE v_0} e^{-i\beta \log(-v_0/\mu)}.
\end{equation}
The integral for $\mathcal{A}(E)$ is a standard result that evaluates to a Gamma function, $|\mathcal{A}(E)|^2 \propto E^{-2} |\Gamma(1 - i \beta)|^2$. Using the identity $|\Gamma(1\pm ix)|^2 = \pi x / \sinh(\pi x)$, the differential particle number spectrum takes its final form:
\begin{equation}
\frac{\dd N}{\dd E} \propto \frac{1}{E^2} \frac{\pi\beta}{\sinh(\pi\beta)}.
\end{equation}
The physics is entirely encoded in the dependence of the phase coefficient $\beta$. For the abelian case, $\lambda \to q$, the $\beta$ is a constant with respect to both charge of the scalar (fixed by the field theory) and the energy. For gravity, in sharp contrast, $\beta\propto E$, which yields the familiar Planck-like spectrum. 

Gravity's thermal form requires two ingredients: the logarithmic phase itself, which is a universal consequence of the long-range potential in four dimensions, and the linear energy dependence of its coefficient. This gravitational energy dependence has a spectacularly simple origin in the double-copy construction of the 3-point vertex: the YM color factor is replaced by a second copy of the kinematic numerator, $\mathcal{A}_{GR} \propto (k\cdot p)^2/r$, effectively promoting the interaction's momentum dependence from linear to quadratic.

Can the thermal nature of gravity and all the mysteries therein truly stem from something as elementary as a tree-level substitution in the three-point amplitude for minimal coupling? Yes, because this thermal structure is already present in the gauge theory, encoded in the spectrum of the color eigenvalue $\lambda$.

 The Yang-Mills amplitude for emitting a particle with a specific color eigenvalue $\lambda$ and energy $E$ has a squared magnitude
\begin{equation}
 |\mathcal{A}(E, \lambda)|^2 \propto \frac{1}{E^2} \frac{\pi C \lambda}{\sinh(\pi C \lambda)},
\end{equation}
where the $1/E^2$ prefactor is characteristic of Bremsstrahlung, but the $\sinh$ term introduces a thermal dependence on the color charge $\lambda$.

\section{The Spectrum of Radiated Color}
 To compute an inclusive spectrum, we must integrate over all available color channels. In the large $N_c$ limit~\footnote{A crucial feature of the color-kinematics duality~\cite{Bern:2008qj} is that the kinematic algebra is universal, holding for any number of external particles. The color algebra of any specific, finite-rank gauge group is not; for any finite $N_c$, certain color factors vanish for specific channels (e.g., the four-point singlet in SU(2)). For the duality to hold universally, the kinematic algebra cannot be restricted to any particular SU($N_c$). Instead, it is dual to a generic algebraic structure satisfying only antisymmetry and the Jacobi identity. The large $N_c$ limit of SU($N_c$) provides one ideal realization of such a structure. Our motivation for this limit should not be taken as any desire to access the planar sector or to simplify color topology. Rather, it is to treat the color space as a stand-in for this generic algebra, ensuring a dense spectrum of non-vanishing color channels.}, the canonical framework for this density of states is Random Matrix Theory~\cite{Anninos:2020ccj,wigner1993characteristic}, which predicts the Wigner semicircle law:
\begin{equation}
 \rho(\lambda) \propto \sqrt{R^2 - \lambda^2},
\end{equation}
where $R$ defines the radius of the eigenvalue distribution. The final differential spectrum of radiated color charge is then the product of the dynamical emission probability and this density of available states, integrated over all radiated energies:
\begin{equation}\label{eq:main_result}
 \frac{\dd N}{\dd \lambda} \propto \underbrace{\frac{C \lambda}{\sinh(\pi C \lambda)}}_{\text{Dynamical Factor}} \times \underbrace{\sqrt{R^2 - \lambda^2}}_{\text{Phase Space Factor}}.
\end{equation}

\section{Analysis and Interpretation}
\label{sec:analysis}

\begin{figure}[h]
 \centering
 \includegraphics[width=\columnwidth]{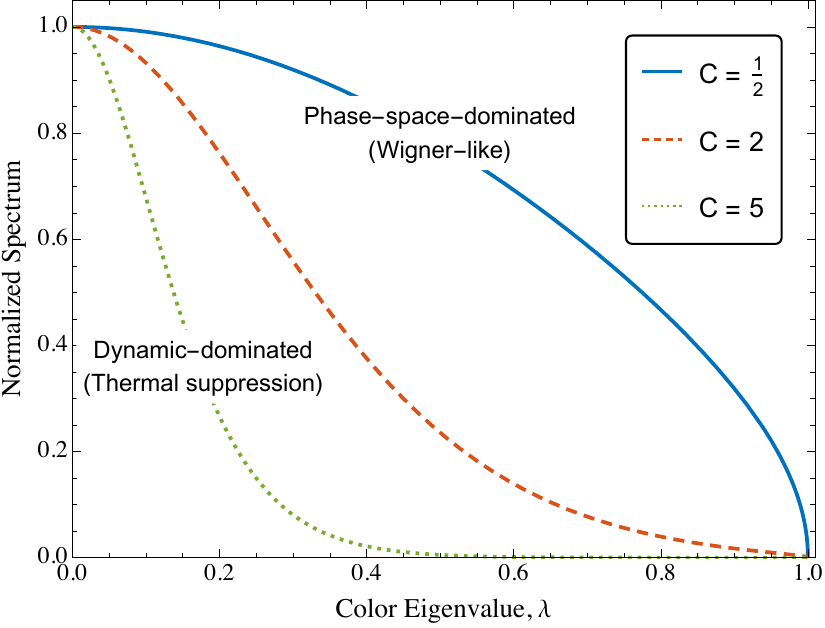} 
 \caption{The spectrum of radiated color charge, $\dd N/\dd \lambda$, for fixed radius $R=1$ and varying effective coupling $C$. As $C$ increases, the spectrum transitions from the phase-space-dominated Wigner semicircle to a dynamically-dominated thermal distribution.}
 \label{fig:spectrum}
\end{figure}

Equation~\eqref{eq:main_result} reveals a competition between dynamics and the available phase space in shaping the radiation.
\begin{itemize}
 \item The \textbf{dynamical factor} is thermal, arising from the universal resummation of soft radiation from a $1/r$ potential. It exponentially suppresses the emission of particles with large color charge ($\lambda \gg 1/C$).
 \item The \textbf{phase space factor} is the Wigner semicircle, the spectral density of color charge. It is maximal at the origin and exhibits algebraic (square--root) suppression toward the boundary ($\lambda \to R$).
\end{itemize}
The observable spectrum's shape depends on the dimensionless ratio $C/R$. For weak coupling ($C \ll R$), the spectrum traces the Wigner semicircle. For strong coupling ($C \gg R$), the thermal factor dominates, producing a Planck-like spectrum truncated by the phase space boundary at $\lambda = R$, as shown in Fig.~\ref{fig:spectrum}. We emphasize that strong coupling in the eikonal context refers to the semi-classical regime of a large eikonal phase where $C\lambda \gg 1$, not a breakdown of perturbation theory; the soft series is already resummed (c.f.~ WKB).

While we have adopted the Wigner semicircle as a well-motivated model~\footnote{The Wigner semicircle nicely reflects the universal eigenvalue density of random Hermitian matrices with SU($N_c$) symmetry and captures the expected large-$N_c$ distribution of color eigenvalues independent of microscopic details. } for the kinematic phase space, we should emphasize that the emergence of thermality does not depend on this specific choice. Rather, the thermal behavior arises from the separate dynamical factor, which we derived from first principles. The thermal dynamics arising from the double-copy thus stands in opposition to the density of states, independent of its precise form. This opposition --- where the strong exponential suppression from the dynamics overwhelms the gentler algebraic fall off of the phase space --- in the large Q regime is a novel physical prediction of the non-abelian root.  The phase-space dominated regime ($C\ll R$) simply corresponds to the perturbative limit, $\chi\ll 1$, where thermal suppression is negligible. The duality holds in both limits, but the specific feature of color thermal dominance is the signature that the gauge theory has entered the non-perturbative regime dual to a semi-classical black hole.

\section{Conclusion}
The double copy provides a rigorous map from non-abelian gauge theory to gravity, uniting classical solutions with quantum radiation. Our analysis of this map has uncovered a novel phenomenon in gauge theory phenomenology: a mechanism for color thermalization driven by coherent sources. We have shown that the radiation from a classical color source, while Bremsstrahlung-like in energy, is thermal in its internal color degrees of freedom. 

This thermal in color-space spectrum suggests that coherent classical non-abelian sources act as ``color scramblers'': generating maximum color-entropy distributions in the charge sector characterized by an effective color temperature $T_C \propto 1/Q$. This offers a clear physical utility on the gauge side: it predicts a thermodynamic regime of high-charge densities where the suppression of emission of high-eigenvalue modes is governed not by kinematics, but by the non-linear structure of the gauge group itself.  

We emphasize that double-copy represents a weak-weak duality.  There is a sharp mapping from gauge states to gravity states.  There is a  sharp mapping of quantum gauge operators to quantum gravitational operators.  This mapping need not mean similar evolution.  In the double-copy web of theories, gauge theories serve as privileged vantage points where we get to see the interplay of their still mysterious universal kinematic algebra against the rigid backbones of their dual gauge color algebras.  Gravity theories on the other hand lack this rigidity with two copies of this kinematic algebra at play together.  In this sense Kerr-Schild and root-Kerr-Schild classical backgrounds provide a Rosetta stone where we can line up structural behavior beyond graph-by-graph analysis as we have demonstrated here. 

Beyond the discovery of a new gauge theory phenomenon, our calculation serves as a stepping stone towards a broader program probing the unitarity of black hole evolution from the perspective of double-copy dual gauge theories. We have identified here that the gross statistical properties of the gravitational field are the double copy of the gross statistical properties of the color field. The underlying gauge evolution while possibly complicated, is manifestly unitary --- simply a rotation of the color frame.  One expects to see the re-emergence of unitarity as correlation in higher order corrections. How protected is the structural relationship at higher orders?  This awaits calculation, but we are optimistic as the symmetries of these systems are very constraining.

As such, we expect the double copy to provide a robust functional arena to sharpen the provocative link between the semi-classical bathing of a probe in soft radiation and the one-loop instability of the quantum vacuum.

\begin{acknowledgments}
We would like to especially thank Dionysius Anninos, Alexander C.~Edison, Juna A.~Kollmeier, Nicolas H.~Pavao, and Sai Sasank Chava for insightful and supportive discussions, as well as comments on an earlier draft. We are particularly grateful for the warm support and encouragement of Donal O'Connell and Chris White who we contacted after ref.~\cite{Aoude:2025jvt} appeared on the arXiv. This work was supported by the DOE under contract DE-SC0015910, and by Northwestern University via the
Amplitudes and Insight Group, Department of Physics and Astronomy, and
Weinberg College of Arts and Sciences.
\end{acknowledgments}

\bibliography{Refs_rootThermality.bib}

\begin{thebibliography}{19}%
\makeatletter
\providecommand \@ifxundefined [1]{%
 \@ifx{#1\undefined}
}%
\providecommand \@ifnum [1]{%
 \ifnum #1\expandafter \@firstoftwo
 \else \expandafter \@secondoftwo
 \fi
}%
\providecommand \@ifx [1]{%
 \ifx #1\expandafter \@firstoftwo
 \else \expandafter \@secondoftwo
 \fi
}%
\providecommand \natexlab [1]{#1}%
\providecommand \enquote  [1]{``#1''}%
\providecommand \bibnamefont  [1]{#1}%
\providecommand \bibfnamefont [1]{#1}%
\providecommand \citenamefont [1]{#1}%
\providecommand \href@noop [0]{\@secondoftwo}%
\providecommand \href [0]{\begingroup \@sanitize@url \@href}%
\providecommand \@href[1]{\@@startlink{#1}\@@href}%
\providecommand \@@href[1]{\endgroup#1\@@endlink}%
\providecommand \@sanitize@url [0]{\catcode `\\12\catcode `\$12\catcode
  `\&12\catcode `\#12\catcode `\^12\catcode `\_12\catcode `\%12\relax}%
\providecommand \@@startlink[1]{}%
\providecommand \@@endlink[0]{}%
\providecommand \url  [0]{\begingroup\@sanitize@url \@url }%
\providecommand \@url [1]{\endgroup\@href {#1}{\urlprefix }}%
\providecommand \urlprefix  [0]{URL }%
\providecommand \Eprint [0]{\href }%
\providecommand \doibase [0]{https://doi.org/}%
\providecommand \selectlanguage [0]{\@gobble}%
\providecommand \bibinfo  [0]{\@secondoftwo}%
\providecommand \bibfield  [0]{\@secondoftwo}%
\providecommand \translation [1]{[#1]}%
\providecommand \BibitemOpen [0]{}%
\providecommand \bibitemStop [0]{}%
\providecommand \bibitemNoStop [0]{.\EOS\space}%
\providecommand \EOS [0]{\spacefactor3000\relax}%
\providecommand \BibitemShut  [1]{\csname bibitem#1\endcsname}%
\let\auto@bib@innerbib\@empty
%</preamble>
\bibitem [{\citenamefont {Duff}(1973)}]{Duff:1973zz}%
  \BibitemOpen
  \bibfield  {author} {\bibinfo {author} {\bibfnamefont {M.~J.}\ \bibnamefont
  {Duff}},\ }\bibfield  {title} {\bibinfo {title} {{Quantum Tree Graphs and the
  Schwarzschild Solution}},\ }\href {https://doi.org/10.1103/PhysRevD.7.2317}
  {\bibfield  {journal} {\bibinfo  {journal} {Phys. Rev. D}\ }\textbf {\bibinfo
  {volume} {7}},\ \bibinfo {pages} {2317} (\bibinfo {year} {1973})}\BibitemShut
  {NoStop}%
\bibitem [{\citenamefont {Monteiro}\ \emph {et~al.}(2014)\citenamefont
  {Monteiro}, \citenamefont {O'Connell},\ and\ \citenamefont
  {White}}]{Monteiro:2014cda}%
  \BibitemOpen
  \bibfield  {author} {\bibinfo {author} {\bibfnamefont {R.}~\bibnamefont
  {Monteiro}}, \bibinfo {author} {\bibfnamefont {D.}~\bibnamefont
  {O'Connell}},\ and\ \bibinfo {author} {\bibfnamefont {C.~D.}\ \bibnamefont
  {White}},\ }\bibfield  {title} {\bibinfo {title} {{Black holes and the double
  copy}},\ }\href {https://doi.org/10.1007/JHEP12(2014)056} {\bibfield
  {journal} {\bibinfo  {journal} {JHEP}\ }\textbf {\bibinfo {volume} {12}},\
  \bibinfo {pages} {056}},\ \Eprint {https://arxiv.org/abs/1410.0239}
  {arXiv:1410.0239 [hep-th]} \BibitemShut {NoStop}%
\bibitem [{\citenamefont {Luna}\ \emph {et~al.}(2015)\citenamefont {Luna},
  \citenamefont {Monteiro}, \citenamefont {O'Connell},\ and\ \citenamefont
  {White}}]{Luna:2015paa}%
  \BibitemOpen
  \bibfield  {author} {\bibinfo {author} {\bibfnamefont {A.}~\bibnamefont
  {Luna}}, \bibinfo {author} {\bibfnamefont {R.}~\bibnamefont {Monteiro}},
  \bibinfo {author} {\bibfnamefont {D.}~\bibnamefont {O'Connell}},\ and\
  \bibinfo {author} {\bibfnamefont {C.~D.}\ \bibnamefont {White}},\ }\bibfield
  {title} {\bibinfo {title} {{The classical double copy for
  Taub{\textendash}NUT spacetime}},\ }\href
  {https://doi.org/10.1016/j.physletb.2015.09.021} {\bibfield  {journal}
  {\bibinfo  {journal} {Phys. Lett. B}\ }\textbf {\bibinfo {volume} {750}},\
  \bibinfo {pages} {272} (\bibinfo {year} {2015})},\ \Eprint
  {https://arxiv.org/abs/1507.01869} {arXiv:1507.01869 [hep-th]} \BibitemShut
  {NoStop}%
\bibitem [{\citenamefont {White}(2024)}]{White:2024pve}%
  \BibitemOpen
  \bibfield  {author} {\bibinfo {author} {\bibfnamefont {C.~D.}\ \bibnamefont
  {White}},\ }\href {https://doi.org/10.1142/q0457} {\emph {\bibinfo {title}
  {{The Classical Double Copy}}}}\ (\bibinfo  {publisher} {World Scientific},\
  \bibinfo {year} {2024})\BibitemShut {NoStop}%
\bibitem [{\citenamefont {Bern}\ \emph {et~al.}(2008)\citenamefont {Bern},
  \citenamefont {Carrasco},\ and\ \citenamefont {Johansson}}]{Bern:2008qj}%
  \BibitemOpen
  \bibfield  {author} {\bibinfo {author} {\bibfnamefont {Z.}~\bibnamefont
  {Bern}}, \bibinfo {author} {\bibfnamefont {J.~J.~M.}\ \bibnamefont
  {Carrasco}},\ and\ \bibinfo {author} {\bibfnamefont {H.}~\bibnamefont
  {Johansson}},\ }\bibfield  {title} {\bibinfo {title} {{New Relations for
  Gauge-Theory Amplitudes}},\ }\href
  {https://doi.org/10.1103/PhysRevD.78.085011} {\bibfield  {journal} {\bibinfo
  {journal} {Phys. Rev. D}\ }\textbf {\bibinfo {volume} {78}},\ \bibinfo
  {pages} {085011} (\bibinfo {year} {2008})},\ \Eprint
  {https://arxiv.org/abs/0805.3993} {arXiv:0805.3993 [hep-ph]} \BibitemShut
  {NoStop}%
\bibitem [{\citenamefont {Bern}\ \emph {et~al.}(2010)\citenamefont {Bern},
  \citenamefont {Carrasco},\ and\ \citenamefont {Johansson}}]{Bern:2010ue}%
  \BibitemOpen
  \bibfield  {author} {\bibinfo {author} {\bibfnamefont {Z.}~\bibnamefont
  {Bern}}, \bibinfo {author} {\bibfnamefont {J.~J.~M.}\ \bibnamefont
  {Carrasco}},\ and\ \bibinfo {author} {\bibfnamefont {H.}~\bibnamefont
  {Johansson}},\ }\bibfield  {title} {\bibinfo {title} {{Perturbative Quantum
  Gravity as a Double Copy of Gauge Theory}},\ }\href
  {https://doi.org/10.1103/PhysRevLett.105.061602} {\bibfield  {journal}
  {\bibinfo  {journal} {Phys. Rev. Lett.}\ }\textbf {\bibinfo {volume} {105}},\
  \bibinfo {pages} {061602} (\bibinfo {year} {2010})},\ \Eprint
  {https://arxiv.org/abs/1004.0476} {arXiv:1004.0476 [hep-th]} \BibitemShut
  {NoStop}%
\bibitem [{\citenamefont {Bern}\ \emph {et~al.}(2024)\citenamefont {Bern},
  \citenamefont {Carrasco}, \citenamefont {Chiodaroli}, \citenamefont
  {Johansson},\ and\ \citenamefont {Roiban}}]{Bern:2019prr}%
  \BibitemOpen
  \bibfield  {author} {\bibinfo {author} {\bibfnamefont {Z.}~\bibnamefont
  {Bern}}, \bibinfo {author} {\bibfnamefont {J.~J.}\ \bibnamefont {Carrasco}},
  \bibinfo {author} {\bibfnamefont {M.}~\bibnamefont {Chiodaroli}}, \bibinfo
  {author} {\bibfnamefont {H.}~\bibnamefont {Johansson}},\ and\ \bibinfo
  {author} {\bibfnamefont {R.}~\bibnamefont {Roiban}},\ }\bibfield  {title}
  {\bibinfo {title} {{The duality between color and kinematics and its
  applications}},\ }\href {https://doi.org/10.1088/1751-8121/ad5fd0} {\bibfield
   {journal} {\bibinfo  {journal} {J. Phys. A}\ }\textbf {\bibinfo {volume}
  {57}},\ \bibinfo {pages} {333002} (\bibinfo {year} {2024})},\ \Eprint
  {https://arxiv.org/abs/1909.01358} {arXiv:1909.01358 [hep-th]} \BibitemShut
  {NoStop}%
\bibitem [{\citenamefont {Adamo}\ \emph {et~al.}(2022)\citenamefont {Adamo},
  \citenamefont {Carrasco}, \citenamefont {Carrillo-Gonz\'alez}, \citenamefont
  {Chiodaroli}, \citenamefont {Elvang}, \citenamefont {Johansson},
  \citenamefont {O'Connell}, \citenamefont {Roiban},\ and\ \citenamefont
  {Schlotterer}}]{Adamo:2022dcm}%
  \BibitemOpen
  \bibfield  {author} {\bibinfo {author} {\bibfnamefont {T.}~\bibnamefont
  {Adamo}}, \bibinfo {author} {\bibfnamefont {J.~J.~M.}\ \bibnamefont
  {Carrasco}}, \bibinfo {author} {\bibfnamefont {M.}~\bibnamefont
  {Carrillo-Gonz\'alez}}, \bibinfo {author} {\bibfnamefont {M.}~\bibnamefont
  {Chiodaroli}}, \bibinfo {author} {\bibfnamefont {H.}~\bibnamefont {Elvang}},
  \bibinfo {author} {\bibfnamefont {H.}~\bibnamefont {Johansson}}, \bibinfo
  {author} {\bibfnamefont {D.}~\bibnamefont {O'Connell}}, \bibinfo {author}
  {\bibfnamefont {R.}~\bibnamefont {Roiban}},\ and\ \bibinfo {author}
  {\bibfnamefont {O.}~\bibnamefont {Schlotterer}},\ }\bibfield  {title}
  {\bibinfo {title} {{Snowmass White Paper: the Double Copy and its
  Applications}},\ }in\ \href@noop {} {\emph {\bibinfo {booktitle} {{Snowmass
  2021}}}}\ (\bibinfo {year} {2022})\ \Eprint
  {https://arxiv.org/abs/2204.06547} {arXiv:2204.06547 [hep-th]} \BibitemShut
  {NoStop}%
\bibitem [{\citenamefont {Bern}\ and\ \citenamefont
  {Grant}(1999)}]{Bern:1999ji}%
  \BibitemOpen
  \bibfield  {author} {\bibinfo {author} {\bibfnamefont {Z.}~\bibnamefont
  {Bern}}\ and\ \bibinfo {author} {\bibfnamefont {A.~K.}\ \bibnamefont
  {Grant}},\ }\bibfield  {title} {\bibinfo {title} {{Perturbative gravity from
  QCD amplitudes}},\ }\href {https://doi.org/10.1016/S0370-2693(99)00524-9}
  {\bibfield  {journal} {\bibinfo  {journal} {Phys. Lett. B}\ }\textbf
  {\bibinfo {volume} {457}},\ \bibinfo {pages} {23} (\bibinfo {year} {1999})},\
  \Eprint {https://arxiv.org/abs/hep-th/9904026} {arXiv:hep-th/9904026}
  \BibitemShut {NoStop}%
\bibitem [{\citenamefont {Carrasco}\ and\ \citenamefont
  {Zekioglu}(2025)}]{Carrasco:2025ymt}%
  \BibitemOpen
  \bibfield  {author} {\bibinfo {author} {\bibfnamefont {J.~J.~M.}\
  \bibnamefont {Carrasco}}\ and\ \bibinfo {author} {\bibfnamefont
  {S.}~\bibnamefont {Zekioglu}},\ }\bibfield  {title} {\bibinfo {title} {{The
  double copy effective action: a quantum (chromodynamics) approach to
  space-time}},\ }\href@noop {} {\  (\bibinfo {year} {2025})},\ \Eprint
  {https://arxiv.org/abs/2511.01799} {arXiv:2511.01799 [hep-ph]} \BibitemShut
  {NoStop}%
\bibitem [{\citenamefont {Aoude}\ \emph {et~al.}(2024)\citenamefont {Aoude},
  \citenamefont {O'Connell},\ and\ \citenamefont {Sergola}}]{Aoude:2024sve}%
  \BibitemOpen
  \bibfield  {author} {\bibinfo {author} {\bibfnamefont {R.}~\bibnamefont
  {Aoude}}, \bibinfo {author} {\bibfnamefont {D.}~\bibnamefont {O'Connell}},\
  and\ \bibinfo {author} {\bibfnamefont {M.}~\bibnamefont {Sergola}},\
  }\bibfield  {title} {\bibinfo {title} {{Amplitudes for Hawking Radiation}},\
  }\href@noop {} {\  (\bibinfo {year} {2024})},\ \Eprint
  {https://arxiv.org/abs/2412.05267} {arXiv:2412.05267 [hep-th]} \BibitemShut
  {NoStop}%
\bibitem [{\citenamefont {Hawking}(1975)}]{Hawking:1975vcx}%
  \BibitemOpen
  \bibfield  {author} {\bibinfo {author} {\bibfnamefont {S.~W.}\ \bibnamefont
  {Hawking}},\ }\bibfield  {title} {\bibinfo {title} {{Particle Creation by
  Black Holes}},\ }\href {https://doi.org/10.1007/BF02345020} {\bibfield
  {journal} {\bibinfo  {journal} {Commun. Math. Phys.}\ }\textbf {\bibinfo
  {volume} {43}},\ \bibinfo {pages} {199} (\bibinfo {year} {1975})},\ \bibinfo
  {note} {[Erratum: Commun.Math.Phys. 46, 206 (1976)]}\BibitemShut {NoStop}%
\bibitem [{Note1()}]{Note1}%
  \BibitemOpen
  \bibinfo {note} {The eikonal limit describes the dynamics of particles whose
  trajectories are nearly light-like and forward-scattered. These are precisely
  the kinematics of the modes that just skim the forming event horizon to
  become the observed Hawking radiation.}\BibitemShut {Stop}%
\bibitem [{\citenamefont {Aoude}\ \emph {et~al.}(2025)\citenamefont {Aoude},
  \citenamefont {O'Connell}, \citenamefont {Sergola},\ and\ \citenamefont
  {White}}]{Aoude:2025jvt}%
  \BibitemOpen
  \bibfield  {author} {\bibinfo {author} {\bibfnamefont {R.}~\bibnamefont
  {Aoude}}, \bibinfo {author} {\bibfnamefont {D.}~\bibnamefont {O'Connell}},
  \bibinfo {author} {\bibfnamefont {M.}~\bibnamefont {Sergola}},\ and\ \bibinfo
  {author} {\bibfnamefont {C.~D.}\ \bibnamefont {White}},\ }\bibfield  {title}
  {\bibinfo {title} {{Hawking Radiation meets the Double Copy}},\ }\href@noop
  {} {\  (\bibinfo {year} {2025})},\ \Eprint {https://arxiv.org/abs/2510.25866}
  {arXiv:2510.25866 [hep-th]} \BibitemShut {NoStop}%
\bibitem [{\citenamefont {Ilderton}\ \emph {et~al.}(2025)\citenamefont
  {Ilderton}, \citenamefont {Lindved},\ and\ \citenamefont
  {Rajeev}}]{Ilderton:2025aql}%
  \BibitemOpen
  \bibfield  {author} {\bibinfo {author} {\bibfnamefont {A.}~\bibnamefont
  {Ilderton}}, \bibinfo {author} {\bibfnamefont {W.}~\bibnamefont {Lindved}},\
  and\ \bibinfo {author} {\bibfnamefont {K.}~\bibnamefont {Rajeev}},\
  }\bibfield  {title} {\bibinfo {title} {{Hawking radiation from the double
  copy}},\ }\href@noop {} {\  (\bibinfo {year} {2025})},\ \Eprint
  {https://arxiv.org/abs/2510.25852} {arXiv:2510.25852 [hep-th]} \BibitemShut
  {NoStop}%
\bibitem [{Note2()}]{Note2}%
  \BibitemOpen
  \bibinfo {note} {A crucial feature of the color-kinematics duality~\cite
  {Bern:2008qj} is that the kinematic algebra is universal, holding for any
  number of external particles. The color algebra of any specific, finite-rank
  gauge group is not; for any finite $N_c$, certain color factors vanish for
  specific channels (e.g., the four-point singlet in SU(2)). For the duality to
  hold universally, the kinematic algebra cannot be restricted to any
  particular SU($N_c$). Instead, it is dual to a generic algebraic structure
  satisfying only antisymmetry and the Jacobi identity. The large $N_c$ limit
  of SU($N_c$) provides one ideal realization of such a structure. Our
  motivation for this limit should not be taken as any desire to access the
  planar sector or to simplify color topology. Rather, it is to treat the color
  space as a stand-in for this generic algebra, ensuring a dense spectrum of
  non-vanishing color channels.}\BibitemShut {Stop}%
\bibitem [{\citenamefont {Anninos}\ and\ \citenamefont
  {M{\"u}hlmann}(2020)}]{Anninos:2020ccj}%
  \BibitemOpen
  \bibfield  {author} {\bibinfo {author} {\bibfnamefont {D.}~\bibnamefont
  {Anninos}}\ and\ \bibinfo {author} {\bibfnamefont {B.}~\bibnamefont
  {M{\"u}hlmann}},\ }\bibfield  {title} {\bibinfo {title} {{Notes on matrix
  models (matrix musings)}},\ }\href {https://doi.org/10.1088/1742-5468/aba499}
  {\bibfield  {journal} {\bibinfo  {journal} {J. Stat. Mech.}\ }\textbf
  {\bibinfo {volume} {2008}},\ \bibinfo {pages} {083109} (\bibinfo {year}
  {2020})},\ \Eprint {https://arxiv.org/abs/2004.01171} {arXiv:2004.01171
  [hep-th]} \BibitemShut {NoStop}%
\bibitem [{\citenamefont {Wigner}(1993)}]{wigner1993characteristic}%
  \BibitemOpen
  \bibfield  {author} {\bibinfo {author} {\bibfnamefont {E.~P.}\ \bibnamefont
  {Wigner}},\ }\bibfield  {title} {\bibinfo {title} {Characteristic vectors of
  bordered matrices with infinite dimensions i},\ }in\ \href@noop {} {\emph
  {\bibinfo {booktitle} {The Collected Works of Eugene Paul Wigner: Part A: The
  Scientific Papers}}}\ (\bibinfo  {publisher} {Springer},\ \bibinfo {year}
  {1993})\ pp.\ \bibinfo {pages} {524--540}\BibitemShut {NoStop}%
\bibitem [{Note3()}]{Note3}%
  \BibitemOpen
  \bibinfo {note} {The Wigner semicircle nicely reflects the universal
  eigenvalue density of random Hermitian matrices with SU($N_c$) symmetry and
  captures the expected large-$N_c$ distribution of color eigenvalues
  independent of microscopic details.}\BibitemShut {Stop}%
\end{thebibliography}%

\end{document}